\documentclass{acm_proc_article-sp}

\usepackage{color}
\usepackage{url}
\usepackage{epsfig}
\usepackage{graphicx}
\usepackage{subfigure}
\usepackage{subfloat}

\begin{document}

\title{Temporal Analysis of Activity Patterns of Editors in Collaborative Mapping Project of OpenStreetMap}

\numberofauthors{3}
\author{
    \alignauthor
    Taha Yasseri\\
           \affaddr{Oxford Internet Institute}\\
           \affaddr{University of Oxford}\\
           \affaddr{Oxford, UK}\\
           \email{taha.yasseri@oii.ox.ac.uk}
           \and
    \alignauthor
    Giovanni Quattrone\\
           \affaddr{Dept. of Computer Science}\\
           \affaddr{University College London}\\
           \affaddr{London WC1E 6BT, UK}\\
           \email{g.quattrone@cs.ucl.ac.uk}
    \and 
    \alignauthor Afra Mashhadi\\
           \affaddr{Bell Labs, Alcatel-Lucent}\\
           \affaddr{Dublin, Ireland}\\
           \email{afra.mashhadi@alcatel-lucent.com}
}

\maketitle

\begin{abstract}
In the recent years Wikis have become an attractive platform for social studies of the human behaviour. Containing millions  records of  edits across the globe, collaborative systems  such as Wikipedia have allowed researchers to gain a better understanding of editors participation and their activity patterns.  However, contributions made to Geo-wikis \textemdash wiki-based collaborative mapping projects\textemdash differ from systems such as Wikipedia in a 
fundamental way due to spatial dimension of the content that limits the contributors to a set of those who posses local knowledge about a specific area and therefore cross-platform studies and comparisons are required to build a comprehensive image of online open collaboration phenomena. In this work, we study the temporal behavioural pattern of OpenStreetMap editors, a successful example of geo-wiki, for two European capital cities. We categorise different type of temporal patterns and report on the historical trend within a period of 7 years of the project age. We also draw a comparison with the previously observed editing activity patterns of Wikipedia.
\end{abstract}




\section{Introduction}\label{intro}

In the recent years, the research on wikis and open, online collaboration systems has attracted a great deal of attention from social scientist, aiming to analyse  behaviour of a large number of individuals and the interaction between them in a collaborative environment in details. The outcome of such research not only could provide a deeper understanding on human collaboration and its intrinsic features in general, but also could have great applications in innovating and improving new and existing collaboration platforms. A common finding has shown that in many of collaborative systems such as Wikipedia, the distribution of editors and contributions is highly skewed, with the majority of contributions coming from only about 2.5\% of contributors~\cite{panciera2009wikipedians}. Wikipedia as a predominant example of mass collaboration has been studied vastly and from multiple aspects, e.g., size and growth rate~\cite{voss2005,almeida2007}, conflict and editorial wars~\cite{sumi2011b,yasseri2012b}, user reputation and article quality~\cite{wilkinson2007,wu2011}, linguistic features and readability~\cite{yasseri2012c}. 

However, despite the great popularity that geo-wiki systems received in the last years, they have seen far less research attention, with the exception of recent studies on accuracy and coverage of OpenStreetMap\footnote{\small \tt http://www.openstreetmap.org/} where the accuracy has been shown to be high~\cite{haklay2010many,mashhadiwiki} and coverage to be non-uniformly distributed~\cite{zielstra2010comparative,mashhadicscw}. However, users behaviour in this domain is by far  less investigated subject and thus requires more research effort to be well understood. In this regard, Hristova et al~\cite{desiICWSM} has taken  the first steps  to quantitatively understand users participation in geo-wikis. The authors show that in OpenStreetMap the distribution of editors and contributions are as skewed as Wikipedia, with 95\% of contributions made in the area of London, UK, attributed to less than 10\% of users. Also Panciera et al~\cite{panciera2010geowiki} found that in Cyclopath\footnote{\small \tt http://cyclopath.org/} 5\% of its users are responsible for the majority of its content. These results are even more interesting when one considers that geo-wikis systems are fundamentally different from classical web-based wikis as  their spatial dimension limits the contributors to a set of those who posses local knowledge about a specific area (i.e., have physically visited the place). For this reason, not all the properties investigated on classical web-based wikis could be assumed to  hold for geo-wikis too.

Furthermore, the recent popularity and widespread adoption of smart-phones has brought forward classical wiki systems as a candidate worth considering in the urban domain too, with citizens becoming cartographers, with geo-wikis such as  OpenStreetMap. Indeed, these services offer mobile applications that users can deploy directly on their smart-phones to generate geo-tagged content {\em on the go}. Given these characteristics, the question we then ask ourselves is: \emph{what are the temporal activity patterns of geo-wiki editors and how they differ from those of previously highlighted in Wikipedia?} We seek answers to this question to not only discover  interesting patterns in human dynamics, but also to learn the extent to which  the rise in use of mobile devices has changed  the work-flow of collaborative mapping projects. The effects and potentials of transformation towards  ``mobile crowdsourcing''  could be well examined by studying characteristics of such platforms in details,  with great implications in other areas related to ubiquitous and social computing.  In this direction, this paper takes the first steps  in identifying the \emph{temporal} behavioural pattern of geo-wiki editors. In particular, it focuses on OpenStreetMap (OSM), a very successful example of geo-wikis which boasts over one million users collectively building a free, openly accessible, editable map of the world.

 Temporal  patterns of human behaviour has been studied extensively in various fields. In the domain of web and networking, the circadian patterns of the Internet traffic have brought interesting information about individual habits of the Internet usage in different societies, assisting with utilisation of infrastructure~\cite{spennemann2007sessional,mocaches,spennemann2006internet}. In domain of social networking and collaborative systems, various works have addressed the \emph{burstiness} nature of user's behaviour both in terms of voluntarily contributions~\cite{yasseri2012b} and edge creations between social network users~\cite{gaito2012bursty}. Finally, Eagle et al.~\cite{eagle2009eigenbehaviors} have studied the temporal behavioural pattern of human by monitoring communication, location and interactions of 100 test subjects. They extracted common ``eigenbehaviour" structures from their data by using principle component analysis.

In this paper we adopt similar methodology and analyse  the activity logs of OSM editors for two European capital cities of London and Rome for a period of 7 and 5 years respectively. We follow the   principal component  analysis to identify the repeating temporal structures of editors in our data, where typical structures are represented by eigen-patterns. Finally, we study these behavioural structures  over different stages of OSM. Our results show that despite the differences between the nature of contributions of geo-wikis and wikis, the OSM editors generally exhibit similar temporal behavioural pattern of contributions during the day as Wikipedians.  Furthermore, we show that over the past years  the editors behaviour has shifted towards afternoon (3 p.m. to 6 p.m.) activities from the late night activities perhaps corresponding to the success of OSM as a smart-phone application. However, this latter has a slower dynamics in the case of Rome compared to London.

\section{Materials and Methods}\label{sec:meth}

\paragraph{Data}

The OSM dataset is freely available to download  and contains the history of all edits (since 2005) on all spatial objects performed by all users. Spatial objects can be one of three types:  {\em nodes},  {\em ways}, and {\em relations}. Nodes broadly refer to POIs, ways are representative of roads, and relations are used for grouping other objects together. For the purpose of this study, we selected two cities to analyse: London, UK, as an example of a very well represented large metropolitan city in OSM, and Rome, Italy, as an example of a large city which is steadily increasing its spatial representation in OSM. Furthermore, we focused only on nodes; in fact, we expect that editing pattern of nodes differ from those found in classical wiki systems due to simplicity of adding/editing a node on the fly in a more ubiquitous fashion. Finally, to consider only genuine users' contributions, we have looked into contributions that most likely correspond to bulk imports. Many bulk imports were detected in the whole dataset, with tens of thousands of edits done in a single day by a single user, spread throughout Greater London (e.g., more than 20,000 post boxes spread across all Greater London appeared in OSM in only one day in 2009 from the same user). We chose to discard such data for the reason that we intend to model genuine `bottom-up', user-generated contributions, of which massive imports are not representative.
The datasets we are left with are summarised below:
\begin{center}
    {\small
        \begin{tabular}{c|ccc}
        \em City & \em \#Users &  \em \#Edits & \em Time Window \\
        \hline
        London & 607   &   534361 & 7 years\\
        Rome &  563  & 146000  & 5 years\\
        \end{tabular}\label{tab:osm}
    }
\end{center}
\paragraph{Method}

To capture the editing patterns of editors in OSM, we compute two different measures:

\begin{itemize}

\item {\em Circadian cycles.} We analyzed the periodic characteristics of temporal patterns of editors for  each hour of 24 hours day by considering the ratio of the average number of edits that has been done within the considered time window on the average number of edits that has been done within all the 24 hours of the day, as specified below:
    $$
        circadian_{\Delta t} = \frac{avg(\#Edits_{\Delta t})}{avg(\#Edits)}
    $$
    where, $avg(\#Edits_{\Delta t})$ is average amount of edits occurred during the specific time interval $\Delta t$ of the day and $avg(\#Edits)$ is the average number of edits occurred during the whole day. 

\item {\em Eigenbehaviours.} We used the principle components analysis (PCA) of our data  focusing on \emph{time of the day}. The principle component analysis has been extensively used in the past to identify the behavioural patterns of users mobility \cite{eagle2009eigenbehaviors,Park2010eigenmode}. Similarly, we adopt the same methodology in which we define  the principal components as a set of vectors that span a `behaviour space' and characterise the behavioural variation between users.  More precisely, we create a  matrix $M$ where each row corresponds to a unique user in our dataset and each column corresponds to the time unit under study. Each entry $m_{ij}$ of our matrix $M$ contains the average ratio of edits that the user $u_i$ has made during the time units $t_j$.  Exactly as before, as time unit we considered one hour time interval. For each column of this matrix, we compute the empirical mean and the deviation of each column from this mean, after this we computed the covariance matrix $B$ of the previous matrix. Finally, we extract the repeating behaviour from this data structure by calculating the eigenvectors and eigenvalues of $B$. These extracted eigenbehaviours are the heavily weighted vectors that represent a type of repeated behaviour, such as editing during evening, night or afternoon. In order to get a robust set of eigenvectors, we needed to filter out editors with
few edits leading to abnormal circadian patterns with picks of larger than 20\% of all edits within a one hour window.
\end{itemize}

\section{Results}\label{sec:res}

\begin{figure}[!h]
    \centering
    \includegraphics[width=0.41\textwidth]{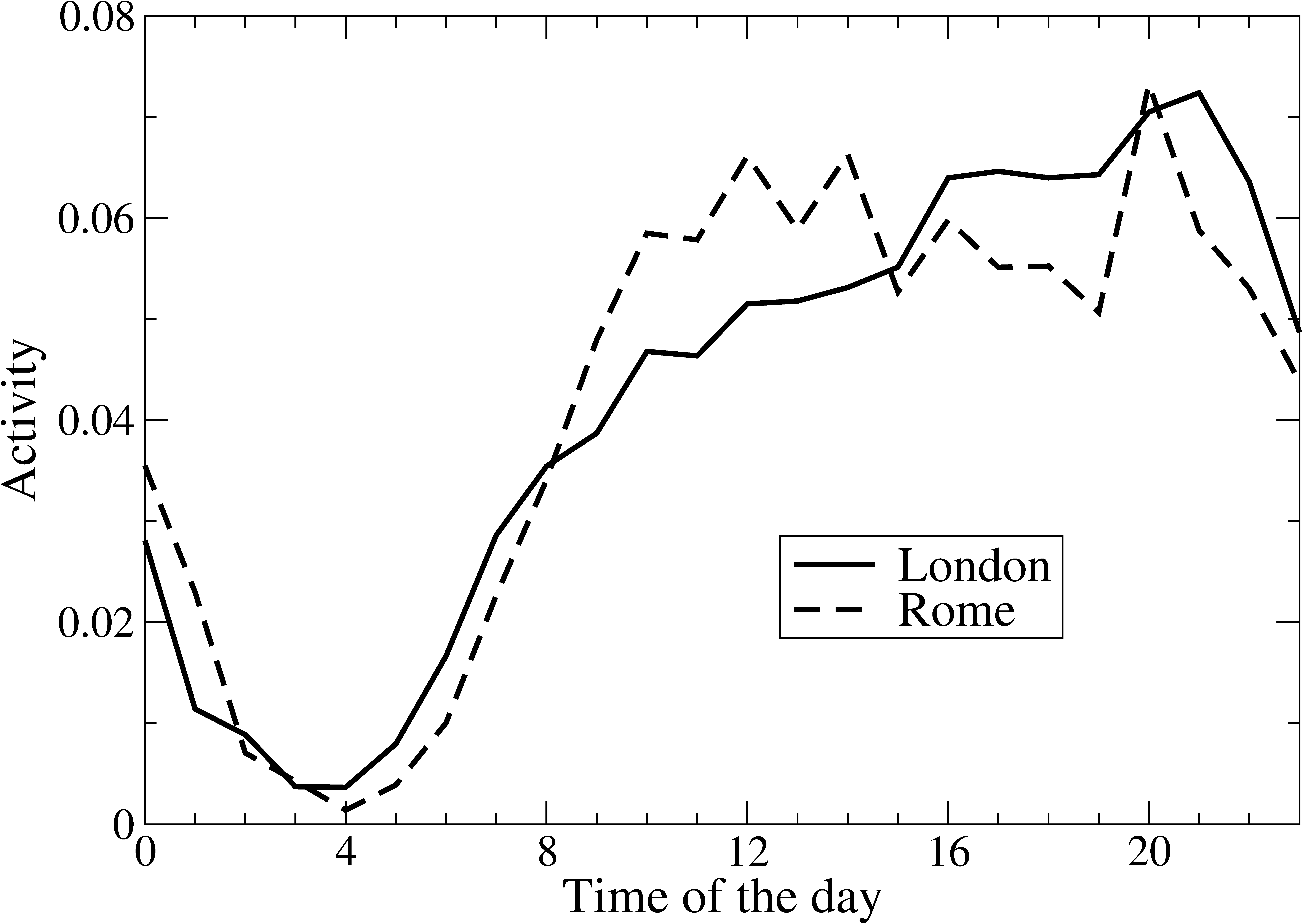}
    \caption{The circadian pattern of edits based on the aggregated data for the whole dataset period.}
    \label{fig:circidianAll}
\end{figure}

We first start by analysing the periodic characteristics of temporal patterns of editors inferring the circadian cycles of OSM edits.  We considered every single edit performed since 2005/2007 on 
London/Rome and used the timestamps assigned to each edit to calculated the overall activity of users for the time of day. Figure~\ref{fig:circidianAll} illustrates the circadian pattern. It is interesting to observe that the editing activities are to their minimum  at early morning around 4 a.m., followed by a rapid increase up to noon. The activity shows a slight increase until around 9 p.m., where it start to decrease during night. This result corresponds closely to  the circadian patterns previously observed in Wikipedia~\cite{yasseri2012circadian} as well as qualitatively corresponding to the other kind of human activities such as mobile calling behaviour~\cite{jo2012circadian}, and the Internet instant messaging~\cite{pozdnoukhov2010exploratory}. However, Rome editors show more activity around noon and less in the afternoon compared to London editors.

\begin{figure}[!h]
    \centering
    \includegraphics[width=0.45\textwidth]{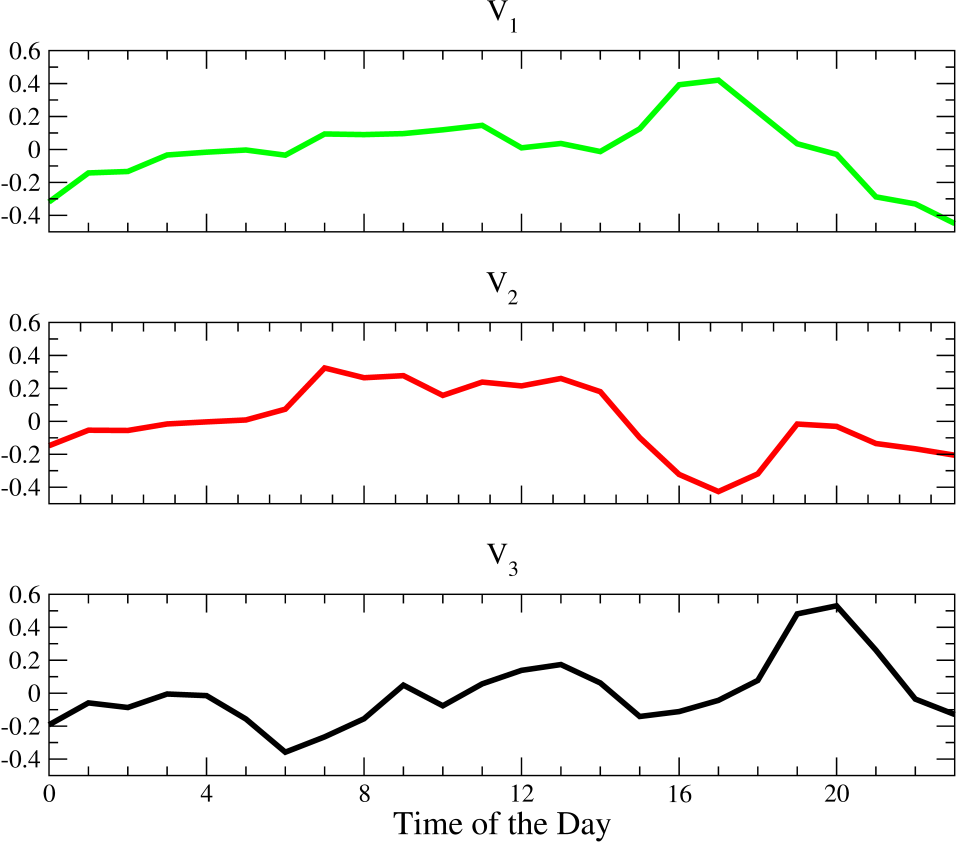}
    \caption{Three Top Eigenbehaviours constructed from individual activity patterns of London OSM editors.}
    \label{fig:eigen}
\end{figure}

We further investigate the circadian activity patterns by performing PCA. The first 3 principal components ${V_1,V_2,V_3}$ extracted from the individual activity pattern of London editors are shown in Figure~\ref{fig:eigen}. The sum of the 3 corresponding eigenvalues exceeds 80\% of the sum of the all spectrum, justifying the reduction of the space dimension from 24 to 3. It is possible to see that  these 3 component roughly correspond to 3 type of editorial pattern: {\em (i)} $V_1$ corresponds to afternoon edits where the peak is from 2 p.m to 6 p.m, {\em (ii)} $V_2$ corresponds to morning and noon edits and finally where the peak of edits falls from 7 a.m to 2 p.m. {\em (iii)} $V_3$ corresponds to night edits where the peak of edits fall from 6 p.m. to midnight. Note that the eigenvectors indicate
the direction of the most typical deviations from the average behaviour. We repeated the same analysis for the Rome data which resulted in a very similar pattern. However as the smaller size of the Rome dataset contributes to the level of noise dramatically, we proceed with our analysis  based on the eigenvectors extracted from London data only.  

\begin{figure}[!h]
    \centering
    \includegraphics[width=0.45\textwidth]{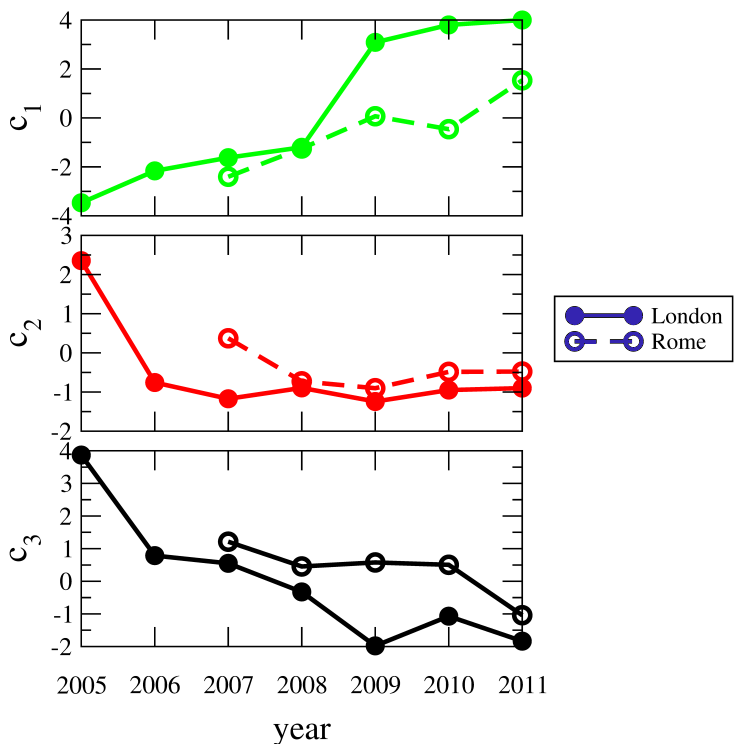}
    \caption{The trend of  projections into the identified  eigenbehaviours (shown in Figure~\ref{fig:eigen}) over years for London and Rome editors.}
    \label{fig:projection}
\end{figure}

In the next step we build up activity vectors for all edits within each year for both London and Rome and project them on the identified eigen-behaviours ${V_1,V_2,V_3}$. The projected values ${c_1,c_2,c_3}$ show to which extent the activity vector could be reproduced by the corresponding eigen-behaviour. $c_i$'s calculated for London and Rome for different years are shown in Figure~\ref{fig:projection}. In both cases a shift towards afternoon edit is evident; however faster in the case of London. Furthermore, we observe a subside effect in the edits associated to night time activities. We speculate two reasons for these trends. Firstly the widespread of Internet services on devices such as smart-phones in the recent years, may have had an impact on the temporal behaviour of users as they become less limited to stationary computers (e.g., at work or home). It is interesting to note that the lesser extend of afternoon projection for Rome can similarly be explained by a recent study\footnote{http://tinyurl.com/ylj9bvb} which has highlighted a much lower adoption rate of smart-phones in  Italy (about 11\%)  in comparison to the UK (70\%) market.  Secondly, as OSM system ages, the saturation effect  on maps may mean that less  mass contributions from power-users is required, thus a decrease in night activities (i.e., a trend usually associated with power-users).

\section{Conclusion}\label{sec:conc}

In this work we studied the temporal behavioural pattern of OpenStreetMap editors as a representative example of geo-wiki systems. By performing Principal Component Analysis, we extracted the typical 
editorial behaviour temporal patterns. We showed that although OSM editors generally exhibit similar temporal behavioural pattern of contributions during the day as Wikipedians, but over the past years the editors behaviour has shifted towards afternoon activities from the late night activities perhaps corresponding to the success of OSM as a smart-phone application. However, this trend has a slower dynamics in the case of Rome compared to London. We believe these results not only shed light on the underlying mechanisms of collaborative mapping projects, but also could suggest elaborative strategies in order
to increase the viability and quality of the project; facilitating more mobile contributions being one example. Note that higher quality of London map compared to Rome has been already reported~\cite{mashhadiwiki}, in accordance with our observation of larger use of mobile devices by its users.  

As future direction, we plan to go beyond this study by considering in our temporal analysis not only the daily pattern of user edits, but also the difference on daily pattern during working days and weekends for different user categories (e.g., power-users, early adopters). Furthermore, to gain a deeper understanding of geo-wikis editors behaviour, we plan to divide our data into roads versus POIs, hypothesising that editing POIs requires less skills and attention than roads, for this reason they may exhibit a different temporal pattern.


\end{document}